\documentclass[preprint,amsmath,amssymb,aps,showkeys,showpacs]{revtex4}
\usepackage[english]{babel}
\usepackage{graphicx}
\usepackage{graphics}
\usepackage{amsmath}
\usepackage{dcolumn}
\usepackage{amssymb}
\usepackage{bm}

\begin{document}
\title{Effects on a Landau-type system for a neutral particle with no permanent electric dipole moment subject to the Kratzer potential in a rotating frame}
\author{Abinael B. Oliveira}
\affiliation{Departamento de F\'isica, Universidade Federal da Para\'iba, Caixa Postal 5008, 58051-900, Jo\~ao Pessoa-PB, Brazil.}

\author{Knut Bakke}
\email{kbakke@fisica.ufpb.br}
\affiliation{Departamento de F\'isica, Universidade Federal da Para\'iba, Caixa Postal 5008, 58051-900, Jo\~ao Pessoa-PB, Brazil.}

\begin{abstract}
The behaviour of a neutral particle (atom, molecule) with an induced electric dipole moment in a region with a uniform effective magnetic field under the influence of the Kratzer potential [A. Kratzer, Z. Phys. {\bf3}, 289 (1920)] and rotating effects is analysed. It is shown that the degeneracy of the Landau-type levels is broken and the angular frequency of the system acquires a new contribution that stems from the rotation effects. Moreover, in the search for bound states solutions, then, it is shown that the possible values of this angular frequency of the system are determined by the quantum numbers associated with the radial modes and the angular momentum, the angular velocity of the rotating frame and by the parameters associated with the Kratzer potential. 

\end{abstract}


\maketitle

\section{Introduction}

Rotating effects has attracted a great deal of discussion in the literature, for instance, Landau and Lifshitz \cite{landau2} pointed out a geometrical point of view where a coordinate transformation from a system at rest to a uniformly rotating frame yields an intriguing behaviour: the line element of the Minkowski spacetime becomes not well-defined for large distances, i.e., the coordinate system becomes singular at large distances. This singular behaviour at large distances is associated with the velocity of the particle would be greater than the velocity of the light. Recently, this spatial constraint has been explored in studies of the Dirac oscillator \cite{b20} and the Landau quantization for neutral particles \cite{b3}. Another context that has investigated rotating effects is in interferometry, where quantum effects associated with the geometric quantum phases has been observed. Examples of these quantum effects are the Sagnac effect \cite{sag,sag2,sag5}, the Mashhoon effect \cite{r4} and the Aharonov-Carmi geometric phase \cite{ac2}. In condensed matter physics, rotating effects have been investigated in Bose-Einstein condensation in ultra cold diluted atomic gases \cite{cond2}, the quantum Hall effect \cite{cond1}, the Aharonov-Carmi geometric phase in C60 molecules \cite{cond3,cond3a}, in quantum rings \cite{r12,r11,dantas} and spintronics \cite{spint1,spint2,spint3}. It is worth mentioning works with Dirac fields \cite{r10}, scalar fields \cite{r8,r8a} and based on the the coupling between the angular momentum and the angular velocity of the rotating frame \cite{r1,r2,r3}.

The aim of this work is to investigate rotating effects on an atom (molecule) with an induced electric dipole moment subject to the Kratzer potential \cite{kratzer,kratzer2,kratzer3} in a region with a uniform effective magnetic field. In recent years, the quantum dynamics of a neutral particle with no permanent electric dipole moment in a region with a uniform effective magnetic field has been made in Ref. \cite{lin3}, where it is shown that an analogue of the Landau quantization \cite{landau} can be observed and it has interests in cold atoms technology \cite{kuklov,pachos,duan,jaksch1,jaksch2,liu,28,29}. In \cite{fur5}, the Landau-type quantization for an atom (molecule) with an induced electric dipole moment was investigated in a two-dimensional quantum ring. In this work, by searching for bound states solutions, we show that the angular frequency of the system differs from the analogue of the cyclotron frequency obtained in Ref. \cite{lin3} and the possible values of this angular frequency of the system are determined by the quantum numbers associated with the radial modes and the angular momentum, the angular velocity of the rotating frame and by the parameters associated with the Kratzer potential \cite{kratzer,kratzer2,kratzer3}.

The structure of this paper is: in section II, we make a brief introduction of the quantum dynamics for a moving atom (molecule) with an induced electric dipole moment and the Landau quantization associated with it; thus, we analyse this Landau-type system subject to the Kratzer potential \cite{kratzer,kratzer2,kratzer3} in a rotating frame; in section III, we present our conclusions.

\section{Rotating effects}

In this section, we investigate rotating effects on an atom (molecule) with an induced electric dipole moment that interacts with external fields subject to the Kratzer potential \cite{kratzer,kratzer2,kratzer3}. Let us begin by introducing the Landau system associated with an atom (molecule) with no permanent electric dipole moment. First of all, in the rest frame of the neutral particle or in the laboratory frame, the electric dipole moment of a moving neutral particle can be considered to be proportional to the electric field: $\vec{d}=\alpha\,\vec{E}$, where $\alpha$ is the dielectric polarizability of the atom (molecule) \cite{whw,griff}. On the other hand, if the neutral particle is moving with a velocity $\vec{v}$ ($v\ll c$), then, it interacts with an electric field $\vec{E}'$ determined by the Lorentz transformation. By applying the Lorentz transformation of the electromagnetic field up to terms of order $\mathcal{O}\left(v^{2}/c^{2}\right)$, we have that the electric field $\vec{E}'$ must be replaced with $\vec{E}'=\vec{E}+\vec{v}\times\vec{B}$, the fields $\vec{E}$ and $\vec{B}$ correspond to the electric and magnetic fields in the laboratory frame, respectively \cite{griff}. Hence, the dielectric polarizability of the atom (molecule) can be written as $\vec{d}=\alpha\left(\vec{E}+\vec{v}\times\vec{B}\right)$ (SI units). Thereby, the Lagrangian of the system must be written in terms of the electric $\vec{E}'=\vec{E}+\vec{v}\times\vec{B}$ as $\mathcal{L}=\frac{1}{2}m\,v^{2}+\alpha\,\vec{v}\cdot\left(\vec{B}\times\vec{E}\right)+\frac{1}{2}\,\alpha\,E^{2}-V$, where $V$ is a scalar potential, $m=M+\alpha\,B^{2}$ is the effective mass of the system and $M$ is the mass of the neutral particle \cite{whw}. Let us work by assuming that $B^{2}=\mathrm{constant}$, hence, the Hamiltonian operator that describes this system is \cite{bf32,ob,lin3,whw} (with the units $c=\hbar=1$):
\begin{eqnarray}
\mathbb{H}_{0}=\frac{\hat{\pi}^{2}}{2m}-\frac{\alpha}{2}\,E^{2}+\mathbb{V},
\label{1.1}
\end{eqnarray}
where $\mathbb{V}$ is the potential energy (scalar operator) and the operator $\hat{\pi}$ is defined as 
\begin{eqnarray}
\hat{\pi}=\hat{p}+\alpha\,\vec{E}\times\vec{B},
\label{1.2}
\end{eqnarray}
where $\hat{p}=-i\vec{\nabla}$ is the momentum operator (vector operator), and the vectors $\vec{E}$ and $\vec{B}$ are the electric and magnetic fields in the laboratory frame, respectively. According to Ref. \cite{whw2}, the term $\alpha\,E^{2}$ given in Eq. (\ref{1.1}) is very small compared with the kinetic energy of the atoms, therefore we can neglect it without loss of generality from now on.

Furthermore, in Ref. \cite{lin3} is shown that a field configuration of crossed magnetic and electric fields can give rise to an analogue of the Landau quantization \cite{landau} for a neutral particle (atom, molecule) with an induced electric dipole moment, where this field configuration is given by  
\begin{eqnarray}
\vec{E}=\frac{\chi\rho}{2}\,\hat{\rho};\,\,\,\,\,\vec{B}=B_{0}\,\hat{z},
\label{1.3}
\end{eqnarray}
where $\rho=\sqrt{x^{2}+y^{2}}$ is the radial coordinate, $\hat{\rho}$ is a unit vector in the radial direction, $\chi$ is a constant related to the uniform volume charge density, $B_{0}$ is a constant and $\hat{z}$ is unit vector in the $z$-direction. It is well-known in the literature that the Landau quantization \cite{landau} takes place when the motion of an electrically charged particle in a plane perpendicular to a uniform magnetic field acquires distinct orbits, and the energy levels of this system becomes discrete and infinitely degenerate. It is important in studies of two-dimensional surfaces \cite{l2,l3,l4}, the quantum Hall effect \cite{l1} and Bose-Einstein condensation \cite{l5,l6}. With this field configuration (\ref{1.3}), thus, we have an effective vector potential given by $\vec{A}_{\mathrm{eff}}=\vec{E}\times\vec{B}=-\frac{\chi\,B_{0}\,\rho}{2}\,\hat{\varphi}$, where $\hat{\varphi}$ is a unit vector in the azimuthal direction. Thereby, by using this effective vector potential, we can determine a uniform effective magnetic field given by $\vec{B}_{\mathrm{eff}}=\vec{\nabla}\times\vec{A}_{\mathrm{eff}}=\vec{\nabla}\times\left(\vec{E}\times\vec{B}\right)=-\chi\,B_{0}\,\hat{z}$. In this sense, a Landau-type system is built for an atom with an induced electric dipole moment that moves in a plane perpendicular to the uniform effective magnetic field $\vec{B}_{\mathrm{eff}}$. It is worth pointing out that a field configuration of crossed electric and magnetic field has attracted interests in studies of hydrogen atom \cite{cross1,cross2,cross3,cross4,cross5,cross6}, large electric dipole moments \cite{cross7}, atoms and molecules in strong magnetic field \cite{cross8,cross9,cross10} and the quasi-Landau behaviour in atomic systems \cite{cross11,cross12}.

Let us now consider the Landau-type system for an atom with no permanent electric dipole moment to be subject to the Kratzer potential \cite{kratzer,kratzer2,kratzer3}:
\begin{eqnarray}
V\left(\rho\right)=-\frac{2D\,a}{\rho}+\frac{D\,a^{2}}{\rho^{2}},
\label{1.3a}
\end{eqnarray}
where $D$ and $a$ are constants. It has attracted a great interest in studies of molecules \cite{kratzer4,molecule,ct5}. Further, if this system rotates with a constant angular velocity $\vec{\Omega}=\Omega\,\hat{z}$, then, it has been discussed in Refs. \cite{landau3,landau4,dantas,anan,r13} that the Hamiltonian operator that describes the behaviour of the system in a rotating frame is given by 
\begin{eqnarray}
\mathbb{H}=\mathbb{H}_{0}-\vec{\Omega}\cdot\hat{L},
\label{1.4}
\end{eqnarray} 
where $\hat{L}$ is the angular momentum operator given by $\hat{L}=\vec{r}\times\hat{\pi}$, where $\hat{\pi}$ is given in Eq. (\ref{1.2}) and $\vec{r}=\rho\,\hat{\rho}$ in a two-dimensional system. In recent decades, geometric quantum phases have been investigated in nonrelativistic quantum systems \cite{anan,r14,r15,r13} based on the approach of Eq. (\ref{1.3}). By substituting Eqs. (\ref{1.1}), (\ref{1.2}) and (\ref{1.3}) into Eq. (\ref{1.4}), the Schr\"odinger equation becomes 
\begin{eqnarray}
i\frac{\partial\psi}{\partial t}&=&-\frac{1}{2m}\left[\frac{\partial^{2}\psi}{\partial\rho^{2}}+\frac{1}{\rho}\frac{\partial\psi}{\partial\rho}+\frac{1}{\rho^{2}}\frac{\partial^{2}\psi}{\partial\varphi^{2}}+\frac{\partial^{2}\psi}{\partial z^{2}}\right]+i\,\frac{\alpha\,\chi\,B_{0}}{2m}\frac{\partial\psi}{\partial\varphi}+\frac{\alpha^{2}\chi^{2}B_{0}^{2}}{8m}\,\rho^{2}\psi\nonumber\\
&+&i\Omega\,\frac{\partial\psi}{\partial\varphi}+\frac{\Omega\,\alpha\,\chi\,B_{0}}{2}\,\rho^{2}\,\psi-\frac{\mu}{\rho}\,\psi+\frac{\tau^{2}}{\rho^{2}}\,\psi,
\label{1.5}
\end{eqnarray}
where $\mu=2\,D\,a$ and $\tau^{2}=D\,a^{2}$.

Note that the Hamiltonian operator of the right-hand-side of Eq. (\ref{1.5}) commutes with the operators $\hat{L}_{z}=-i\,\frac{\partial}{\partial\varphi}$ and $\hat{p}_{z}=-i\,\frac{\partial}{\partial z}$, then, a particular solution to Eq. (\ref{1.5}) can be written in terms of the eigenvalues of  $\hat{L}_{z}$ and  $\hat{p}_{z}$ as $\psi\left(t,\,\rho,\,\varphi,\,z\right)=e^{-i\mathcal{E}t}\,e^{i\,l\,\varphi}\,e^{ikz}\,f\left(\rho\right)$, where $l=0,\pm1,\pm2,\ldots$, $k$ is a constant and $f\left(\rho\right)$ is a function of the radial coordinate. Henceforth, we assume that $k=0$ in order to reduce the system to a planar system. By substituting this particular solution into Eq. (\ref{1.5}) and by performing a change of variables given by $r=\sqrt{m\,\varpi}\,\rho$, hence, the Schr\"odinger equation (\ref{1.5}) becomes
\begin{eqnarray}
\frac{d^{2}f}{dr^{2}}+\frac{1}{r}\frac{df}{dr}-\frac{\gamma^{2}}{r^{2}}\,f-r^{2}\,f+\frac{\vartheta}{r}\,f+\frac{\beta}{m\varpi}\,f=0,
\label{1.8}
\end{eqnarray}
where we have defined the parameters in Eq. (\ref{1.8}):
\begin{eqnarray}
\varpi^{2}&=&\frac{\omega^{2}}{4}+\Omega\,\omega;\nonumber\\
\gamma^{2}&=&l^{2}+2m\tau^{2};\nonumber\\
\beta&=&2m\mathcal{E}+2m\,\Omega\,l+m\omega\,l;\label{1.9}\\
\vartheta&=&\frac{2m\mu}{\sqrt{m\,\varpi}};\nonumber\\
\omega&=&\frac{\alpha\,\chi\,B_{0}}{m}.\nonumber
\end{eqnarray}
It is worth emphasizing that the parameter $\omega$ was defined in Ref. \cite{lin3} as being the cyclotron frequency of the Landau quantization associated with an atom with an induced electric dipole moment.

We proceed with the analysis of the asymptotic behaviour of the possible solutions to Eq. (\ref{1.8}). The asymptotic behaviour is determined for $r\rightarrow0$ and $r\rightarrow\infty$ (singular points), then, it is required that the solution to Eq. (\ref{1.8}) to be finite at $r\rightarrow0$ and $r\rightarrow\infty$. For this purpose, we consider the radial wave function to be well-behaved at $r\rightarrow0$ and vanish at $r\rightarrow\infty$, and thus, we can write the function $f\left(r\right)$ in terms of an unknown function $H\left(r\right)$ as follows \cite{eug,mhv,vercin,heun}:
\begin{eqnarray}
f\left(r\right)=e^{-\frac{r^{2}}{2}}\,r^{\left|\gamma\right|}\,H\left(r\right),
\label{1.10}
\end{eqnarray}
and thus, by substituting Eq. (\ref{1.10}) into Eq. (\ref{1.8}), we have that the function $H\left(r\right)$ is a solution to the following second order differential equation
\begin{eqnarray}
\frac{d^{2}H}{dr^{2}}+\left[\frac{2\left|\gamma\right|+1}{r}-2r\right]\frac{dH}{dr}+\left[\nu+\frac{\vartheta}{r}\right]H=0,
 \label{1.11}
\end{eqnarray}
where $\nu=\frac{\beta}{m\,\varpi}-2-2\left|\gamma\right|$. The second order differential equation (\ref{1.11}) is called as the biconfluent Heun function \cite{heun}, and the function $H\left(r\right)$ is the biconfluent Heun function \cite{heun}: $H\left(r\right)=H_{B}\left(2\left|\gamma\right|,\,0,\,\frac{\beta}{m\,\varpi},\,2\vartheta,\,-r\right)$.

Henceforth, let us use the Frobenius method \cite{arf} in order to write the solution to Eq. (\ref{1.11}) as a power series expansion around the origin: $H\left(r\right)=\sum_{k=0}^{\infty}a_{k}\,r^{k}$. By substituting this series into Eq. (\ref{1.11}), we obtain the recurrence relation:
\begin{eqnarray}
a_{k+2}=-\frac{\vartheta}{\left(k+2\right)\left(k+2+2\left|\gamma\right|\right)}\,a_{k+1}-\frac{\nu-2k}{\left(k+2\right)\left(k+2+2\left|\gamma\right|\right)}\,a_{k},
\label{1.13}
\end{eqnarray}
and $a_{1}=-\frac{\vartheta}{\left(1+2\left|\gamma\right|\right)}\,a_{0}$.

Let us start with $a_{0}=1$, then, from Eq. (\ref{1.13}), we can obtain other coefficients of the power series expansion $H\left(r\right)=\sum_{k=0}^{\infty}a_{k}\,r^{k}$. For instance, the coefficients $a_{1}$, $a_{2}$ and $a_{3}$ are given by
\begin{eqnarray}
a_{1}&=&-\frac{\vartheta}{\left(1+2\left|\gamma\right|\right)};\nonumber\\
a_{2}&=&\frac{\vartheta^{2}}{2\left(2+2\left|\gamma\right|\right)\left(1+2\left|\gamma\right|\right)}-\frac{\nu}{2\left(2+2\left|\gamma\right|\right)};\label{1.14}\\
a_{3}&=&-\frac{\vartheta^{3}}{6\left(3+2\left|\gamma\right|\right)\left(2+2\left|\gamma\right|\right)\left(1+2\left|\gamma\right|\right)}+\frac{\nu\,\vartheta}{6\left(3+2\left|\gamma\right|\right)\left(2+2\left|\gamma\right|\right)}+\frac{\left(\nu-2\right)\vartheta}{3\left(3+2\left|\gamma\right|\right)\left(1+2\left|\gamma\right|\right)}.\nonumber
\end{eqnarray}

We focus on achieving bound states solutions, therefore, we need to impose that the biconfluent Heun series becomes a polynomial of degree $n$. In this way, we guarantee that the function $f\left(r\right)$ becomes well-behaved at the origin and vanishes at $r\rightarrow\infty$.  From the recurrence relation (\ref{1.13}), we have that the biconfluent Heun series becomes a polynomial of degree $n$ by imposing that \cite{bf,ob,eug}:
\begin{eqnarray}
\nu=2n;\,\,\,\,\,a_{n+1}=0,
\label{1.15}
\end{eqnarray}
where $n=1,2,3,\ldots$. By analysing the condition $\nu=2n$, we obtain a general expression for the energy levels
\begin{eqnarray}
\mathcal{E}_{n,\,l}=\sqrt{\frac{\omega^{2}}{4}+\Omega\,\omega}\,\,\left[n+\left|\gamma\right|+1\right]-\frac{1}{2}\,\omega\,l-\Omega\,l,
\label{1.16}
\end{eqnarray}
where $n=1,2,3,\ldots$ is the quantum number associated with the radial modes, $l=0,\pm1,\pm2,\ldots$ is the angular momentum quantum number and $\omega$ is called as the cyclotron frequency of the Landau-type system \cite{lin3} which is defined in Eq. (\ref{1.9}). In particular, we have in Eq. (\ref{1.16}) the coupling between the angular momentum quantum number $l$ and the angular velocity $\Omega$ which corresponds to the Page-Werner {\it et al} term \cite{r1,r2,r3}.

Next, let us analyse the condition $a_{n+1}=0$ given in Eq. (\ref{1.15}). For this purpose, let us consider the cyclotron frequency $\omega$ \cite{lin3} can be adjusted in such a way that the condition $a_{n+1}=0$ can be satisfied. This is possible because we can adjust either the intensity of the magnetic field $B_{0}$ or the intensity of the electric field through the parameter $\chi$ associated with the uniform volume charge density \cite{ob}. With this assumption, we have that both conditions imposed in Eq. (\ref{1.15}) are satisfied and a polynomial solution to the function $H\left(r\right)$ is obtained. As an example, let us take $n=1$ and label $\omega=\omega_{n,\,l}$. For $n=1$ we have the ground state of the system, then $a_{n+1}=a_{2}=0$, and thus the possible values of the cyclotron frequency associated with the ground state of the system are 
\begin{eqnarray}
\omega_{1,\,l}=2\Omega\left[-1\pm\sqrt{1+\frac{4m^{2}\mu^{4}}{\Omega^{2}\left(1+2\left|\gamma\right|\right)}}\right].
\label{1.19}
\end{eqnarray}
Note that the possible values of $\omega_{1,\,l}$ given in Eq. (\ref{1.19}) yields $\varpi>0$, and thus the asymptotic behaviour of the radial wave function when $r\rightarrow\infty$ is satisfied. 

Hence, this example shows us that only specific values of the angular frequency $\omega$ are allowed and depend on the quantum numbers $\left\{n,\,l\right\}$ of the system and the angular velocity of the rotating frame. From the quantum mechanics point of view, this relation of the angular frequency $\omega$ to the quantum numbers of the system $\left\{n,\,l\right\}$ and the the angular velocity of the rotating frame is a quantum effect that stems from the influence of the rotating effects and the Kratzer potential on the Landau-type quantization associated with the neutral particle (atom, molecule) with an induced electric dipole moment. Observe that, by substituting (\ref{1.19}) into Eq. (\ref{1.16}), we have 
\begin{eqnarray}
\mathcal{E}_{1,\,l}=\frac{2m\,\mu^{2}\left(\left|\gamma\right|+2\right)}{\left(1+2\left|\gamma\right|\right)}\mp\Omega\,l\,\sqrt{1+\frac{4m^{2}\mu^{4}}{\Omega^{2}\left(1+2\left|\gamma\right|\right)^{2}}},
\label{1.19a}
\end{eqnarray}
which corresponds to the allowed energies for the ground state of the system. Moreover, the radial wave function (\ref{1.10}) associated with the ground state is given by
\begin{eqnarray}
f_{1,\,l}\left(r\right)=e^{-\frac{r^{2}}{2}}\,r^{\left|\gamma\right|}\left[1-\frac{\vartheta}{\left(1+2\left|\gamma\right|\right)}\,r\right].
\label{1.20}
\end{eqnarray}

Finally, let us rewrite the energy levels (\ref{1.16}) in a general form as
\begin{eqnarray}
\mathcal{E}_{n,\,l}=\sqrt{\frac{\omega^{2}_{n,\,l}}{4}+\Omega\,\omega_{n,\,l}}\,\,\left[n+\left|\gamma\right|+1\right]-\frac{1}{2}\,l\,\omega_{n,\,l}-\Omega\,l.
\label{1.21}
\end{eqnarray}

Hence, the energy levels (\ref{1.21}) are obtained from the effects of the Kratzer potential and rotating effects on the Landau-type system for a neutral particle (atom, molecule) with an induced electric dipole moment. Note that the energy levels are modified in contrast to that obtained in Ref. \cite{lin3} for the Landau quantization, where the ground state of the system becomes determined by the quantum number $n=1$ instead of the quantum number $n=0$, and the degeneracy of the Landau levels is broken as we can see in Eqs. (\ref{1.16}), (\ref{1.19}) and (\ref{1.19a}). By comparing with the cyclotron frequency of the Landau-type quantization $\omega=\frac{\alpha\,\chi\,B_{0}}{m}$ given in Ref. \cite{lin3}, we have an angular frequency given by $\varpi=\sqrt{\frac{\omega^{2}}{4}+\Omega\,\omega}$ which means that the angular frequency of the Landau-type system is modified by the influence of the rotating frame. Besides, only some specific values of the cyclotron frequency $\omega$ are allowed in order that a polynomial solution to the function $H\left(r\right)$ can be obtained, where the allowed values depend on the quantum numbers $\left\{n,\,l\right\}$, the the angular velocity of the rotating frame and the parameters associated with the Kratzer potential as we can see in Eq. (\ref{1.19}) for the ground state of the system. Moreover, observe that the last term of Eq. (\ref{1.21}) corresponds to the coupling between the angular momentum quantum number $l$ and the angular velocity $\Omega$ which is called as the Page-Werner {\it et al} term \cite{r1,r2,r3}. Further, by taking $\Omega\rightarrow0$, the rotating effects vanish, and thus the effects analogous to a Coulomb-type potential effects on the Landau quantization for a neutral particle (atom, molecule) with an induced electric dipole moment are recovered \cite{ob}.

\section{Conclusions}

We have discussed effects of rotation and the Kratzer potential \cite{kratzer,kratzer2,kratzer3} on the Landau quantization associated with neutral particle (atom, molecule) with a induced electric dipole moment. We have seen that the angular frequency of the system acquires a new contribution that stems from the rotation effects. Besides, the Landau-type levels obtained in Ref. \cite{lin3} are modified, where the degeneracy of the Landau-type levels is broken and the ground state of the system becomes determined by the quantum number $n=1$ instead of the quantum number $n=0$. Moreover, a quantum effect characterized by the dependence of the cyclotron frequency of the Landau-type quantization on the quantum numbers of the system and the angular velocity of the rotating frame is obtained, which means that only some specific values of the cyclotron frequency $\omega$ are allowed in order to achieve bound state solutions. A new contribution to the analogue of the Landau levels arises from the coupling between the angular velocity of the rotating frame and the angular momentum, which is called as the Page-Werner {\it et al} term \cite{r1,r2,r3}. Finally, in the limit $\Omega\rightarrow0$, we have that the rotating effects vanish, and thus analogous effects to the Coulomb-type interaction on the Landau-type system are recovered \cite{ob}.

\acknowledgments

The authors would like to thank the Brazilian agencies CNPq and CAPES for financial support.

\end{document}